\documentclass{aa}
\usepackage[version=3]{mhchem}

\usepackage{placeins}
\usepackage{txfonts}
\usepackage{natbib}

\newcommand{\kms}{km s$^{-1}$}
\newcommand{\yb}{YSES~2b}
\newcommand{\mj}{$\mathrm{M_{Jup}}$}
\newcommand{\msol}{$\mathrm{M_\odot}$}
\usepackage{graphicx}

\usepackage{color}
\usepackage{hyperref}
\hypersetup{%
        colorlinks,
        breaklinks=true,
        plainpages=false,%
        citecolor=[rgb]{0,0.20,0.45},
        linkcolor=[rgb]{0,0.20,0.45},
        urlcolor=[rgb]{0,0.20,0.45},
        bookmarksopen=true,%
        bookmarksnumbered=false,%
        bookmarksdepth=5%
}

\usepackage[nolist,nohyperlinks]{acronym}
\newacro{yses}[YSES]{Young Suns Exoplanet Survey}
\newacro{sphere}[SPHERE]{Spectro-Polarimetric High-contrast Exoplanet REsearch}

\makeatletter
\renewcommand*\aa@pageof{, page \thepage{} of \pageref*{LastPage}}
\makeatother

\newcommand{\loD}{\hbox{$\lambda/D$}\xspace}

\begin{document} 

\authorrunning{Kenworthy et al.}
\titlerunning{YSES~2b is a background star}
  \title{YSES~2b is a background star}
  
  \subtitle{Differential astrometric M-dwarf measurements in time}

\author{Matthew Kenworthy\inst{1}
\and
Tomas Stolker\inst{1}
\and
Jens Kammerer\inst{2}
\and
William Balmer\inst{3}
\and
Arthur Vigan\inst{4}
\and
Sylvestre Lacour\inst{5}
\and
Gilles Otten\inst{6}
\and
Eric Mamajek\inst{7}
\and
Christian Ginski\inst{8}
\and
Mathias Nowak\inst{9}
\and
Steven Martos \inst{10}
\and
Jason Wang\inst{11}
\and
Emily Rickman\inst{12}
\and
Markus Janson\inst{13}
\and
Alexander Bohn\inst{1}
\and
Mariangela Bonavita\inst{14}
}

\institute{Leiden Observatory, Leiden University, Postbus 9513, 2300 RA Leiden, The Netherlands
 \and
European Southern Observatory, Karl-Schwarzschild-Straße 2, 85748 Garching, Germany
 \and
Department of Physics \& Astronomy, Johns Hopkins University, 3400 N. Charles Street, Baltimore, MD 21218, USA
 \and
Aix Marseille Univ, CNRS, CNES, LAM, Marseille, France
 \and
LIRA, Observatoire de Paris, Université PSL, Sorbonne Université, Université Paris Cité, CY Cergy Paris Université, CNRS, 92190 Meudon, France
 \and
Academia Sinica, Institute of Astronomy and Astrophysics, 11F Astronomy-Mathematics Building, NTU/AS campus, No. 1, Section 4, Roosevelt Rd., Taipei 10617, Taiwan
 \and
Jet Propulsion Laboratory, California Institute of Technology, 4800 Oak Grove Drive, M/S 321-162, Pasadena, CA, 91109, USA
 \and
School of Natural Sciences, Center for Astronomy, University of Galway, Galway, H91 CF50, Ireland
 \and
LIRA, Observatoire de Paris, Université PSL, Sorbonne Université, Université Paris Cité, CY Cergy Paris Université, CNRS, 92190 Meudon, France
 \and
Univ. Grenoble Alpes, CNRS, IPAG, 38000, Grenoble, France
 \and
Center for Interdisciplinary Exploration and Research in Astrophysics (CIERA) and Department of Physics and Astronomy, Northwestern University, Evanston, IL 60208, USA
 \and
European Space Agency (ESA), ESA Office, Space Telescope Science Institute, 3700 San Martin Drive, Baltimore, MD 21218, USA
 \and
Department of Astronomy, Stockholm University, AlbaNova University Center, 10691 Stockholm, Sweden
 \and
Royal Observatory, Blackford Hill, Edinburgh, EH9 3HJ, Scotland
 }

\date{Received 30 June 2025; accepted 04 August 2025}

  \abstract
   {}
   {We wish to confirm the nature of \yb{}, a purportedly faint companion of the young star YSES~2.}
   {We used on-sky observations from SPHERE and GRAVITY to measure the astrometric position of 2b with respect to the star YSES~2, and examined the competing hypotheses of (i) a bound substellar companion versus (ii) a distant unrelated background source with a non-zero proper motion.}
   {\yb{} appears to be a late-type M-dwarf star over 2 kiloparsecs behind the star YSES~2.
   It has a transverse velocity of $\sim 300$ \kms{} and is located within one of the spiral arms of the Galaxy.
   The main discriminant was multiple epochs of GRAVITY astrometry that identified the sub-milliarcsecond parallactic motion of the star.}
   {}


   \maketitle
%

   \section{Introduction}\label{sec:intro}

The number of directly imaged exoplanets is small enough that each merits significant observational attention.
\yb{} was announced as the third planet discovered as part of the \ac{yses}, a survey of 71 young stars with masses of $\sim$ 1\msol{} in the ($\sim 17$ Myr old) Sco-Cen OB association \citep{Bohn20b}.
The first two planets were discovered around YSES~1 \citep{Bohn20} at separations of 160~au and 320~au, with masses of $14\pm3$ \mj{} and $6\pm1$ \mj{}, respectively.
Their large projected separation from the primary star has made them ideal targets for monitoring and characterisation, most notably with JWST spectroscopy \citep{Hoch25}, which have confirmed their planetary nature.

Candidate companions are distinguished from distant background stars that have similar apparent magnitudes via the common parallax (CPx) test: if the foreground star has a significant proper motion across the sky, then companions that are gravitationally bound to the star share that proper motion, with the vectorial addition of Keplerian orbital motion around the star.
Images taken at two epochs separated enough in time to significantly detect the star's proper motion can then be used to identify sources that appear to be co-moving companions.

The object YSES~2b appears to have a proper motion identical within measured errors to that of the star YSES~2 between two epochs \citep{Bohn21}, and it has colours consistent with a 6 \mj{} planet, based on $H$ and $K_s$ magnitudes and model isochrones from the AMES-COND and AMES-Dusty models \citep{Chabrier00,Allard01}.
A third epoch of observation with the Very Large Telescope (VLT) interferometer GRAVITY showed changes in the relative position of the star and 2b consistent with a Keplerian orbit with modest eccentricity, $e>0.6$, suggesting an active formation pathway for the companion and prompting follow-up observations of the system.

Subsequent astrometric measurements in the intervening five years have proved challenging to interpret, and only with the latest astrometric measurements have we come to a new conclusion: 2b is not an exoplanet orbiting YSES~2 but instead an approximately 2 kiloparsec distant late-type star that is close to the line of sight of the star and has an almost identical proper motion.
YSES~2b now joins the group of candidate companions that were later identified as not being gravitationally bound substellar companions, including CS~Cha~b \citep{Ginski18,Hafffert20} and HD~131399~Ab \citep{Wagner16,Nielsen17}.

We present our observations and re-analysis of imaging data from the Spectro-Polarimetric High-Contrast Exoplanet Research (SPHERE) instrument and GRAVITY observations (Sect.~\ref{sec:obs}), discuss the resultant astrometry and GRAVITY spectrum (Sect.~\ref{sec:results}), and compare the fits assuming Keplerian orbital motion versus a distant background source with non-zero proper motion.
We conclude that YSES~2b is a distant background M dwarf (Sect.~\ref{sec:conc}).

\section{Observations and data reduction}\label{sec:obs}

The earliest astrometric measurements of YSES~2b are detailed in Table~2 of \citet{Bohn21} and the derived astrometric measurements in their Table~3.
Subsequent observations are detailed below.

\subsection{SPHERE}

New coronagraphic SPHERE \citep{Beuzit19} data were obtained on 30 January 2025 as part of a pre-imaging observation in the context of the High-Resolution Imaging and Spectroscopy of Exoplanets \citep[HiRISE; ][]{Vigan2024,Denis25} survey, where accurate astrometry is required to place the single-mode fibre of the instrument at the location of the companions to enable high-spectral resolution characterisation with the CRIRES \citep[CRIRES; ][]{Kaeufl04} spectrograph in the $H$ band.

The data were acquired with the Infra-Red Dual Imaging and Spectrograph \citep[IRDIS; ][]{Dohlen08} in dual-band imaging mode \citep{Vigan2010} with the broadband $K_s$ filter instead of the usual K12 filter pair to maximise the signal-to-noise ratio.
The sequence included a flux calibration of the stellar point spread function off-centred from the coronagraphic mask, a centring imaging with waffle spots, and sky background calibration.
The coronagraphic sequence included 12 images of 64\,s of integration time.
Because the target was observed at relatively high airmass ($\sim$1.3) and the total integration time was short, the field of view rotation is of only 5.7$^\circ$, which results in a motion of $\sim$1.9\,\loD of the point spread function  of YSES~2b throughout the observation.

The previous and new SPHERE data were all reduced using the \texttt{vlt-sphere} automated python pipeline \citep{Vigan2020ascl} using standard calibrations.
Each of the images in the coronagraphic observing sequences are background subtracted and divided by the flat field in the appropriate filters.
Bad pixels are corrected using bad pixel maps created with the official SPHERE ESO pipeline by replacing them with the median of neighbouring good pixels.
All images are corrected from the anamorphic distortion that is known to affect the SPHERE near-infrared data \citep{Maire2021}, and are then aligned to a common centre using the star centre data acquired at the beginning of the coronagraphic sequences.
For this purpose, the four satellite spots inside the adaptive optics control radius are fitted with a 2D Gaussian function.
The accuracy of the centring using this procedure has been determined to be better than 0.1 pixel ($\sim$1.2~mas) for bright stars \citep[S/N $>$ 50 in satellite spots; e.g.][]{Zurlo2014,Zurlo2016}.
For each IRDIS field and filter pair ($H_2$ and $H_3$) taken at other epochs, the calibration process is applied independently to each of the two wavelengths that are acquired simultaneously with IRDIS, resulting in two separate pre-processed angular differential imaging (ADI) data cubes.

The ADI data cubes are processed with the LAM-ADI pipeline \citep{Vigan2015,Vigan2016}.
The two first epoch observations have less than 2$^\circ$ of field-of-view rotation, resulting in major self-subtraction effects for YSES~2b.
For this reason, the images for these datasets are simply de-rotated, median-combined, and a simple spatial-filtering in a box of 5$\times$5\,\loD is used to remove the residual stellar halo and thermal background.
For the other two datasets, we used a principal component analysis implementation following the Karhunen-Lo{\`e}ve image projection  approach \citep{Soummer2012}, with only a single mode subtracted from the images before de-rotation.

YSES~2b is recovered with SPHERE in all filters with a signal-to-noise ratio greater than five.
The precise astrometry and photometry of YSES~2b are estimated using `negative fake companion' subtraction in the pre-processed ADI data cubes \citep{Marois2010}.
A rough estimation of the object position and contrast is first performed using a 2D Gaussian fit.
These initial guesses are then used as a starting point for a Levenberg-Marquardt least-squares minimisation routine, where the position and contrast of the negative fake companion are varied to minimise the residual noise after ADI processing in a circular aperture of radius 1\loD that is centred on the position of the YSES~2b.
When a minimum is reached, the position and contrast of the fake companion are taken as the optimal values for the astrometry and photometry.
The error bars for the fitting process are computed by varying the position and contrast of the fake companion until the variation of the reduced $\chi^{2}$ reaches a level of 1$\sigma$. 

We used the values reported by \citet{Maire2021} for the plate scale and north orientation of the IRDIS images to convert the pixel positions of YSES~2b into sky coordinates.
We did not use a dedicated astrometric calibration for this re-analysis work as subsequent observations with the Very Large Telescope Interferometer (VLTI) have much smaller astrometric errors.

\subsection{GRAVITY}

We observed YSES~2b five times between 20 March 2022 and 2 June 2023 using the VLTI GRAVITY instrument \citep{GravityCollab2017}.
The fringe tracker \citep{Lacour2019} was placed at the location of the host star, and the science fibre at the predicted location of the companion.
Observations were taken in dual field, off-axis mode, which necessitated the subsequent observation of a calibrator binary to correctly phase reference the data \citep{Nowak2024binarycalib}.
On 10 May 2023, after observing the companion, we observed the host star using the science fibre in order to amplitude reference the companion observations.

We reduced the uncalibrated GRAVITY data with the ESO GRAVITY pipeline \citep{Lapeyrere2014} version 1.7.0\footnote{\url{https://www.eso.org/sci/software/pipelines/gravity}} and extracted the companion astrometry from the \texttt{astrored} data products using the \texttt{exogravity}\footnote{\url{https://gitlab.obspm.fr/mnowak/exogravity}} pipeline following previous work \citep{gravity2020,nowak2020}.
The resulting detections are visualised in Fig.~\ref{fig:gravity_detections}.
For the observations on 10 May 2023, which could be amplitude-referenced, the contrast spectrum was extracted and corrected for fibre injection losses.
From each of the four GRAVITY epochs, we also obtained a GRAVITY contrast spectrum at a spectral resolution of $R \sim 500$ in the $K$ band (approximately $2-2.4~\mu m$).
For one of the four GRAVITY epochs, the host star YSES~2 was used as an amplitude reference to flux-calibrate the spectrum, and for the other three, the HD~91881~AB and HD~123227~AB swap (binary star) references were used.
To convert the resulting companion contrast spectra into companion flux spectra, model spectra of the amplitude reference sources are required.

The model spectra of YSES~2, HD~91881~AB, and HD~123227~AB were obtained by fitting a BT-NextGen \citep{allard2012} stellar model atmosphere to archival \textit{Gaia}, Tycho, Two-Micron All Sky Survey \citep[2MASS; ][]{Skrutskie06}, and Wide-field Infrared Survey Explorer \citep[WISE; ][]{Wright10} photometry (and for YSES~2 also the \textit{Gaia} XP spectrum).
For the two binary stars, we additionally used the contrast between A and B measured by GRAVITY and averaged over the $K$ band.
We inferred the best fitting stellar model atmosphere using the \texttt{species} toolkit \citep{stolker2020}, which employs nested sampling with \texttt{PyMultiNest} \citep{feroz2008,buchner2014} to obtain the model parameters with the highest likelihood.
The model atmospheres for the three stellar systems are shown in Appendix~\ref{app:stellar_sed}.
For YSES~2, the surface gravity was set to $\log g = 4.32$~[cgs] and the interstellar extinction was set to $A_V = 0.3279$~mag based on the values reported in \textit{Gaia} Data Release 3 \citep[DR3;][]{gaia2023}.
For HD~91881~AB, we used Gaussian priors of $T_\text{eff,A} = 6117\pm9$~K \citep{gaia2023}, $M_\text{A} = 1.31\pm0.10~\text{M}_\odot$ \citep{tokovinin2014}, $M_\text{B} = 1.07\pm0.10~\text{M}_\odot$ \citep{tokovinin2014}, and $M_\text{B}/M_\text{A} = 0.857\pm0.002$ \citep{makarov2021}.
For HD~123227~AB, we used Gaussian priors of $M_\text{A} = 1.27\pm0.10~\text{M}_\odot$ \citep{tokovinin2014} and $M_\text{B} = 1.21\pm0.10~\text{M}_\odot$ \citep{tokovinin2014}. For both binary stars, the extinction was set to $A_V = 0$~mag.

We combined the four GRAVITY epochs into a single companion flux spectrum by computing a covariance-weighted mean spectrum, noting that the absolute flux calibration of the three epochs using swap (binary star) calibrators as contrast reference are affected by systematics as the companion and the calibrator were observed some time apart.
The YSES~2 SC observation was done directly after the companion observations, providing a better reference for flux calibration.
We scaled the three binary calibrator epochs to best match the flux spectrum obtained from the host star calibrator epoch (10 May 2023).
This was done before computing the covariance-weighted mean spectrum.

\section{Results}\label{sec:results}

\begin{table*}[tp]
    \centering
    \caption{SPHERE-measured astrometry of YSES~2b.}
    \begin{tabular}{ccccccccccccc}
\hline\hline
Date & MJD & Filter & Wavelength & $\delta$RA & $\delta$DEC & $\delta$ mag \\
\hline\hline
2018-04-29 & 58238.063022 & $B_H$ & 1.625 & $-454.55\pm1.43$ & $-953.84\pm1.73$ &  $10.33\pm0.11$ \\
2018-04-29 & 58238.063022 & $B_H$ & 1.625 & $-454.67\pm1.06$ & $-954.66\pm1.29$ &  $10.35\pm0.11$ \\
2020-12-07 & 59191.346270 & $B_{K_s}$ & 2.182 & $-440.86\pm3.76$ & $-953.81\pm5.13$ & $10.01\pm0.21$ \\
2020-12-07 & 59191.346270 & $B_{K_s}$ & 2.182 & $-439.82\pm3.48$ & $-955.98\pm2.58$ & $9.95\pm0.25$ \\
2022-04-10 & 59680.100791 & $D_{H2}$ & 1.593 & $-420.87\pm0.29$ & $-944.04\pm0.42$ & $10.33\pm0.02$ \\
2022-04-10 & 59680.100791 & $D_{H3}$ & 1.667 & $-420.18\pm0.39$ & $-944.71\pm1.99$ & $10.26\pm0.02$ \\
2025-01-30 & 60706.259935 & $B_{K_s}$ & 2.182 & $-404.53\pm2.03$ & $-939.74\pm1.87$ & $10.09\pm0.10$ \\
2025-01-30 & 60706.259935 & $B_{K_s}$ & 2.182 & $-402.49\pm2.19$ & $-931.98\pm3.47$ & $9.95\pm0.07$ \\
    \hline
    \end{tabular}
 
    \label{tab:sphereastrometry}
\end{table*}

\subsection{Astrometric background analysis}

\begin{figure}[tp]
    \begin{centering}
        \includegraphics[width=1.0\linewidth]{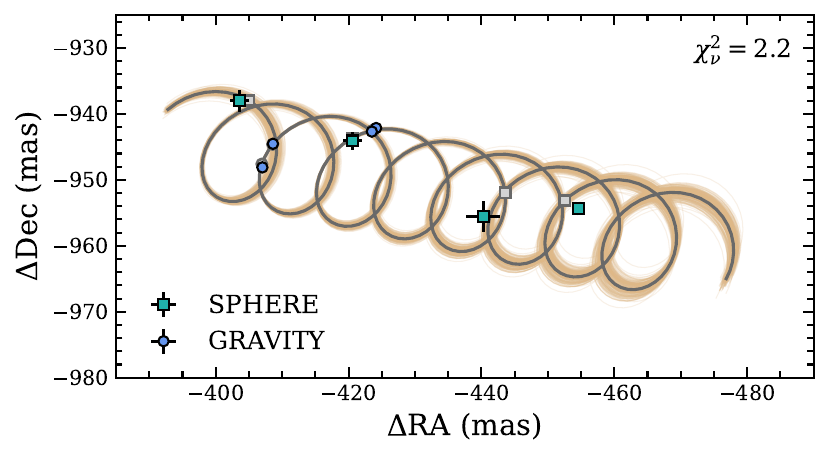}
        \caption{Background fit of the relative astrometry.
        The figure shows 200 background tracks that are randomly drawn from posterior samples.
        The solid grey line is the model calculated from the median parameters.
        The measurements are shown with coloured markers and their respective epochs of the best-fit model as grey markers.}
        \label{fig:background_fit}
    \end{centering}
\end{figure}

The relative astrometry of YSES~2b are reported in Table~\ref{tab:sphereastrometry}.
These measurements are analysed with the non-stationary background model of \texttt{backtracks}\footnote{\url{https://github.com/wbalmer/backtracks}} \citep{backtracks_zenodo}.
This code samples the parallax, coordinates, and proper motions in RA and Dec that generate helical background motion best describing the data. 
The tool optionally uses prior information from \textit{Gaia}, namely the inverse gamma Galactic distance/parallax prior from \citet{BailerJones2021}, and priors on the local proper motions of stars based on a query of nearby \textit{Gaia} sources. 
The parameters of the background object were sampled with the static nested sampling algorithm in the \texttt{dynesty} package \citep{Speagle2020}, with 500 live points and an acceptance fraction of 0.05.

\begin{figure}[tp]
    \begin{centering}
        \includegraphics[width=1.0\columnwidth]{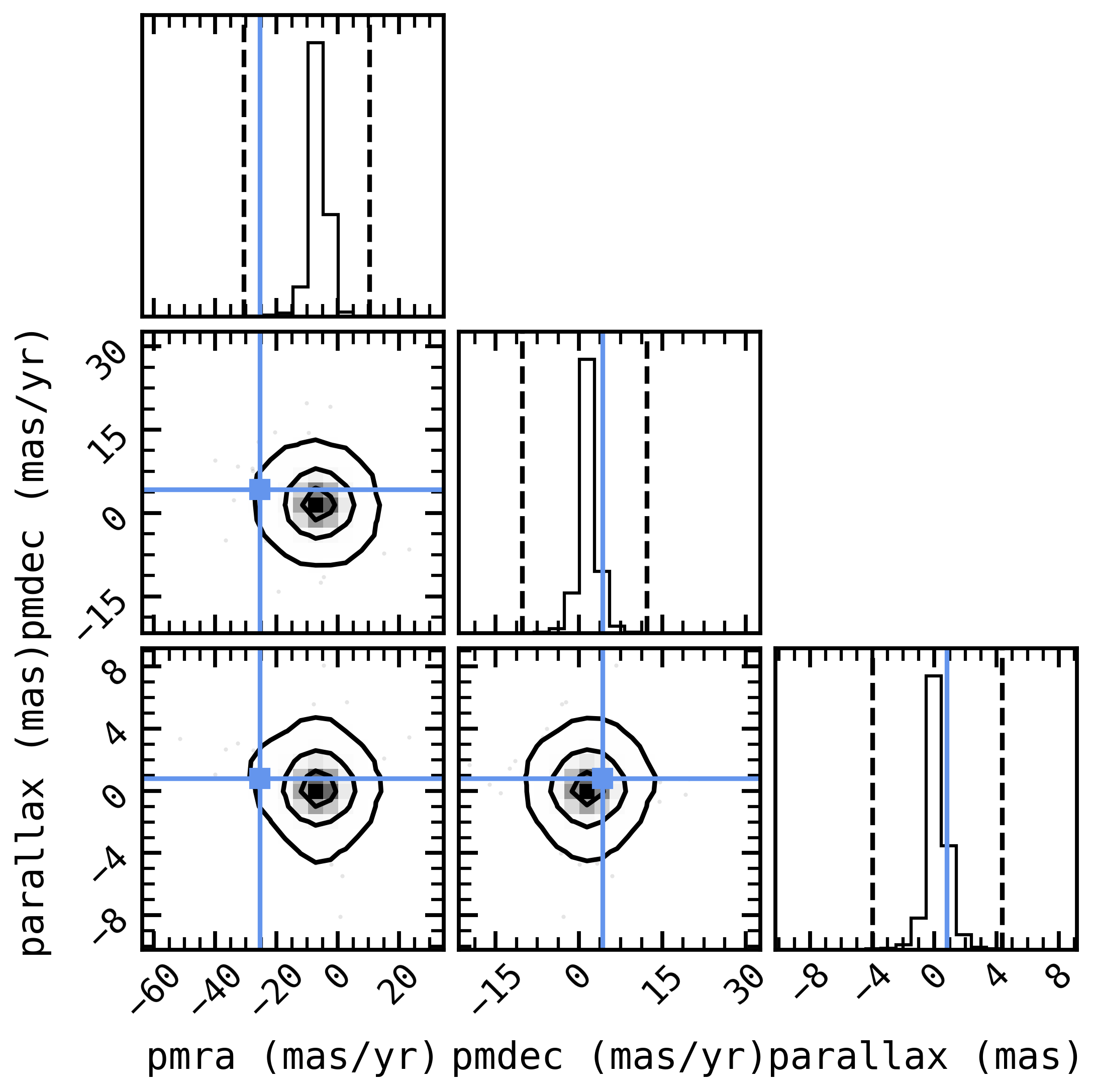}
        \caption{Comparison with the proper motions and parallaxes of all \textit{Gaia} sources within 0.2~deg of YSES~2.
        The inferred parameters of the background star are indicated by the blue dots and lines.}
        \label{fig:pm_background_comparison}
    \end{centering}
\end{figure}

\subsection{Spectral analysis}



\begin{figure}[tp]
    \begin{centering}
        \includegraphics[width=1.0\linewidth]{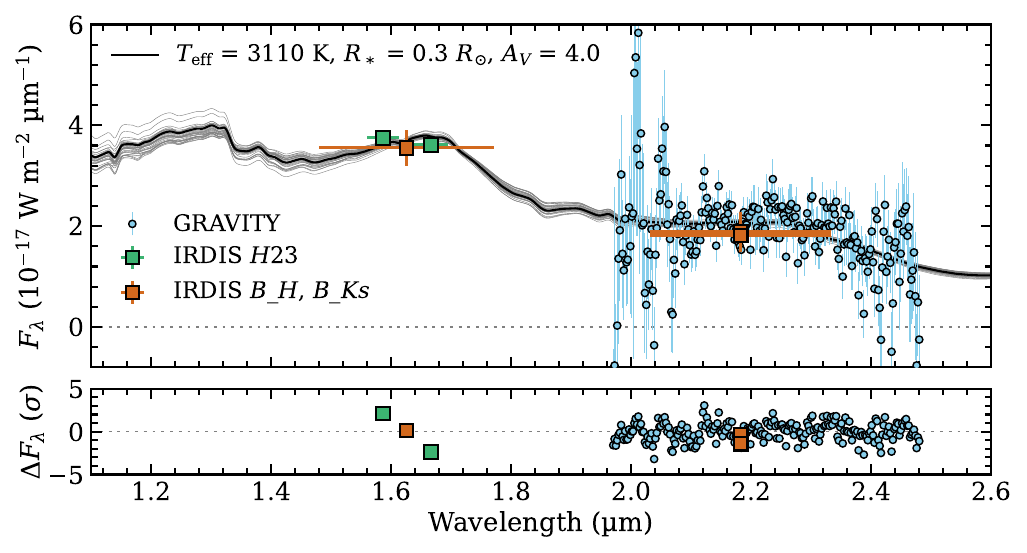}
        \caption{Near-infrared SED of YSES~2b.
        The best-fit model spectrum is shown as black lines, and the grey lines are randomly drawn spectra from the posterior distribution.
        The residuals in the lower panel are normalised by the uncertainties of the data.}
        \label{fig:sed}
    \end{centering}
\end{figure}

The SPHERE contrast measurements are converted into fluxes by calculating synthetic photometry from the model spectrum of the star that was also used for calibrating the GRAVITY spectrum.
We computed magnitudes of 8.51, 8.44, 8.51, and 8.39 in the IRDIS $H2$, $H3$, $B_H$, and $B_{K_s}$ filter, respectively.
We then modelled the available photometry and spectrum with the \texttt{species} toolkit by using the BT-Settl grid of synthetic spectra \citep{allard2012}.
We adopted a normal prior on the parallax based on the posterior from the background analysis.

Figure~\ref{fig:sed} shows a comparison of the data with model spectra.
The retrieved stellar parameters are $T_\mathrm{eff} = 3110 \pm 100$~K, $\log\,g = 4.8 \pm 0.4$, $R_\ast = 0.28 \pm 0.04~\text{R}_\odot$, and $A_V = 4.1 \pm 0.3$.
The constraint on the extinction is in particular driven by the $H$-band photometry: to improve the determination of $A_V$ would require shorter-wavelength data, where the extinction is more significant.
The stellar parameters do however seem reasonable when comparing with parameters retrieved from the PARSEC evolutionary model \citep{nguyen2022}.
For comparison, the bolometric luminosity of YSES~2b, $\log L/L_\odot = -2.17 \pm 0.12$, is consistent with a low-mass star, $M_\ast = 0.3~\text{M}_\odot$, with $T_\mathrm{eff} = 3150$~K, $\log\,g = 5.0$, and $R_\ast = 0.3~\text{R}_\odot$, when assuming an age of 5~Gyr and adopting the parallax from the background fit.

\section{Discussion}\label{sec:discuss}

\subsection{Number of observations required}

Two epochs can confirm the common proper motion of two sources on the sky within the precision of the measurements and, with a long enough time baseline, distinguish them from distant stationary background objects.
For objects that are physically unrelated to a foreground source, more than two epochs are required, to allow for the extra degrees of freedom from the proper motion of the background object (which may not necessarily be identical to the foreground motion) and the distance of the background source.

The $\chi^2_\nu$ of a model to the data provides a measure of the goodness of this fit.
We took the 8 epochs of astrometry of YSES~2b and, starting with the first three epochs, proceeded to fit a background model to the astrometry with increasing numbers of astrometric measurements, to determine the goodness of fit as a function of number of observations.
The final fit of {\tt backtracks} to the eight epochs are shows in Fig.~\ref{fig:background_fit} with a $\chi^2_\nu$ of 3.02, and the corner plot for the derived parameters of the background object are shown in Fig.~\ref{fig:backtrack_corner}.

\subsection{The nature of the background object}




The parallax of the background object is $0.41^{+0.37}_{-0.28}$ mas, with a proper motion of pmRA=$-25.42^{+0.25}_{-0.18}$ mas yr$^{-1}$ and pmDec=$4.40^{+0.28}_{-0.30}$ mas yr$^{-1}$.
The spectral energy distribution (SED) of the background object, together with an SED fit from stellar models, is shown in Fig.~\ref{fig:sed}.
The best fit to the SED is a $T=3065$~K star, with a $\log g$ of 4.4 and radius of 0.5 $\text{R}_\odot$, with an optical extinction of 2.7.
The galactic latitude and longitude of the star is $l=300^\circ, d=+3.1^\circ$.
Combined with the derived distance of 2.5 kpc, this places the star within the Sagittarius arm of the Galaxy, and the SED is consistent with an M-dwarf star with extinction consistent with this path length.

A 2D histogram of all the proper motions within a one degree radius circle centred on YSES~2 (Fig.~\ref{fig:pm_background_comparison}) shows that the proper motion of this M dwarf is 3$\sigma$ away from the mean of the distribution of proper motions in that region of the sky. 
The coincidence is even more unfortunate, as this considers only the magnitude of the proper motion and not its direction: a significantly different direction on the sky would have identified it as a background source much earlier.
The projected transverse velocity at a distance of 2.5~kpc corresponds to $\sim 300$ \kms{}.

\subsection{Bound versus background}

Common proper motion is the classical benchmark test for determining physical companionship in direct imaging studies.
However, as the case of YSES-2b demonstrates, the result of such a test by itself is not always conclusive, since some background stars can move deceptively similarly to the target stars.
This is particularly relevant for relatively distant associations such as Sco-Cen, where the proper motion of the targets stars can be as low as $\sim$20~mas yr$^{-1}$.
Colour criteria can help in the candidate vetting but comes with its own shortcomings, such as reliance on theoretical SED models that may not be representative of all real planets and that distant reddened stars can have similar colours.

This highlights the merit of CPx versus non-common parallax (NCPx) as a test to distinguish ambiguous planet candidates.
In pathological cases where a background star moves with a very similar speed and direction on the sky as the target star, reaching a robust conclusion about companionship based on common proper motion could conceivably take years or decades, even with a high astrometric precision.
By contrast, as long as the target star is sufficiently nearby, CPx versus NCPx can provide robust companionship testing over timescales of as little as months.
For a case such as YSES~2, where the parallax is 9.15 mas (109 pc) and the GRAVITY astrometric precision is $\sim$0.1 mas, CPx can be confirmed or refuted to within a relative distance of $\sim$3.6 pc (at 3$\sigma$) between the target star and the candidate companion.
A positive CPx test would thus firmly exclude any possible background stars, which typically reside at distances of thousands of parsecs. 

Tests of CPx can be performed as part of astrometric campaigns for other purposes, such as common proper motion  testing or orbital monitoring, but the optimal time sampling is somewhat distinct.
For CPx testing, it is of central importance that (some of) the astrometric data be acquired at different parts of the year, since it is the phase difference in the Earth's orbit that sets the parallactic baseline.
The ideal scheduling for such testing is pairing of points at $\pm$3 months relative to the conjunction of the star -- in other words, typically near the start and end of the target's yearly observability window.

\section{Conclusions}\label{sec:conc}

The direct imaging of exoplanets remains challenging, especially when following up on candidate companions with astrometric measurements precise enough to distinguish Keplerian motion from background stars with non-zero proper motion.
GRAVITY is an excellent instrument that can identify these background objects clearly and rapidly with an appropriate observing cadence.
The impact of extremely large telescope instruments, with their corresponding increase in astrometric precision due to the larger primary mirror diameter, will resolve this far more quickly in future observations.

Single candidate companions at large projected separations are the most challenging to confirm with astrometry, especially if the expected orbital motion is comparable to the parallax of the star or a background source.
Ironically, multiple exoplanets in a system are less susceptible to this, as the probability of having more than one background source with the same confounding proper motion on the sky is significantly less likely than for just one source.
Acceleration consistent with Keplerian motion around the stellar component(s) provides the strongest evidence for a bound companion, followed by spectroscopic data that are consistent with other empirical spectra from other confirmed bound companions or evolutionary models for low-mass objects.
Long-term monitoring with GRAVITY+, and the higher spatial resolution of the extremely large telescopes, will enable rapid differentiation between distant background objects and gravitationally bound substellar companions.

\section*{Data availability}    

An online repository with materials used in this work is available at \url{https://github.com/mkenworthy/2b_or_not_2b} using the {\tt showyourwork!} package \citep{Luger2021}.

\begin{acknowledgements}

Based on observations collected at the European Southern Observatory under ESO programmes 108.226K.002 and 1104.C-0651 (ExoGRAVITY Large programme).
This research has used the SIMBAD database, operated at CDS, Strasbourg, France \citep{wenger2000}.
This research has made use of the Jean-Marie Mariotti Center \texttt{Aspro} service \footnote{Available at http://www.jmmc.fr/aspro}.
This work has used data from the European Space Agency (ESA) mission {\it Gaia} (\url{https://www.cosmos.esa.int/gaia}), processed by the {\it Gaia} Data Processing and Analysis Consortium (DPAC, \url{https://www.cosmos.esa.int/web/gaia/dpac/consortium}).
Funding for the DPAC has been provided by national institutions, in particular the institutions participating in the {\it Gaia} Multilateral Agreement.
To achieve the scientific results presented in this article we made use of the \emph{Python} programming language\footnote{Python Software Foundation, \url{https://www.python.org/}}, especially the \emph{SciPy} \citep{virtanen2020}, \emph{NumPy} \citep{numpy}, \emph{Matplotlib} \citep{Matplotlib}, \emph{emcee} \citep{foreman-mackey2013}, \emph{astropy} \citep{astropy_1,astropy_2}, \emph{PynPoint} \citep{Amara2012PYNPOINTexoplanets,Stolker2019PynPointdata} packages.
AV acknowledges support from Agence Nationale de la Recherche (ANR) under grant ANR-23-CE31-0006-01 (MIRAGES) and from the European Research Council (ERC) under the European Union’s Horizon 2020 research and innovation programme, grant agreement No. 757561 (HiRISE).
S.L.\ acknowledges the support of the French Agence Nationale de la Recherche (ANR-21-CE31-0017, ExoVLTI) and of the European Research Council (ERC Advanced Grant No. 101142746, PLANETES).
J.J.W. is supported by NASA XRP Grant 80NSSC23K0280 and the Alfred P. Sloan Foundation.
We thank the referee for their careful reading of this manuscript.

\end{acknowledgements}

\bibliographystyle{aa}
\bibliography{bib}

\begin{appendix}

\section{Stellar SED fits}
\label{app:stellar_sed}

\begin{figure}[h!]
    \begin{centering}
        \includegraphics[width=\columnwidth]{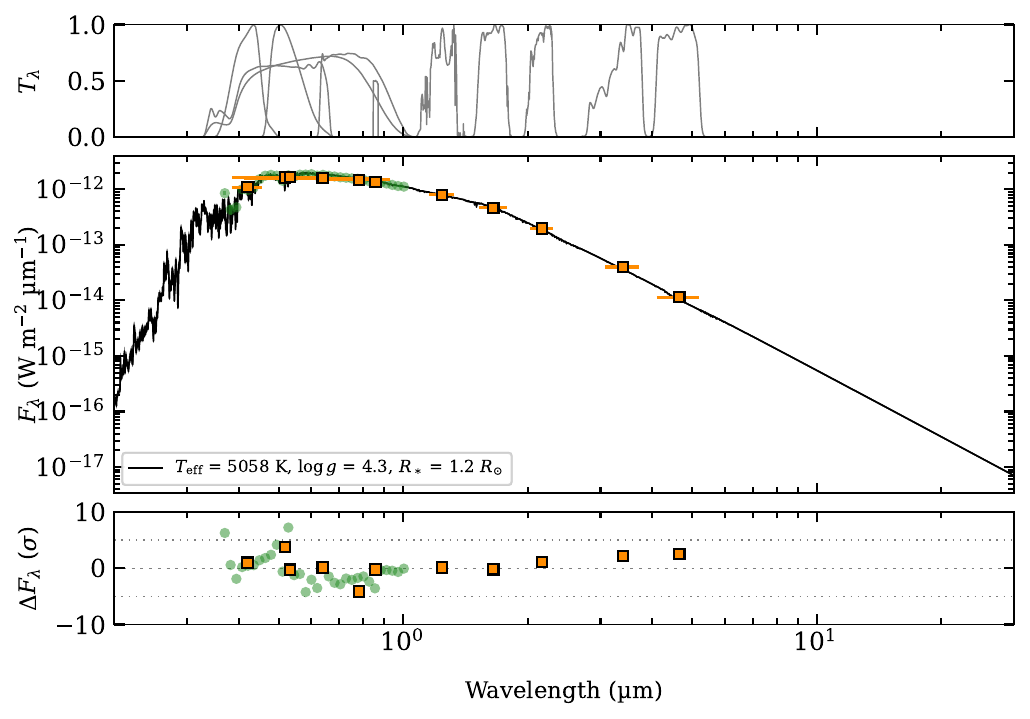}
        \includegraphics[width=\columnwidth]{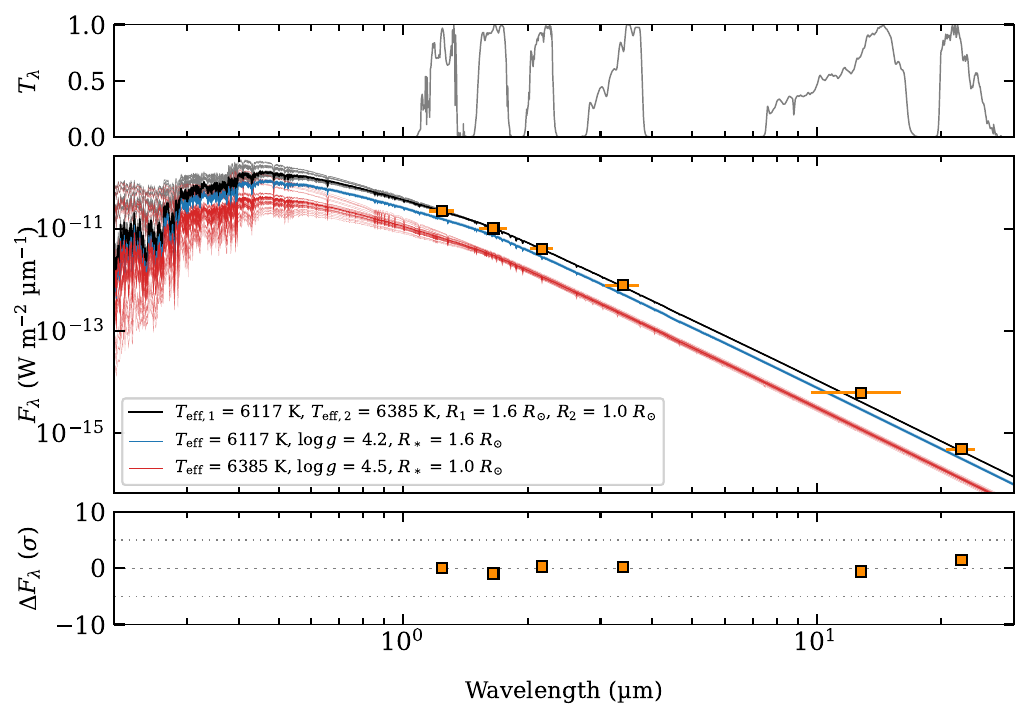}
        \includegraphics[width=\columnwidth]{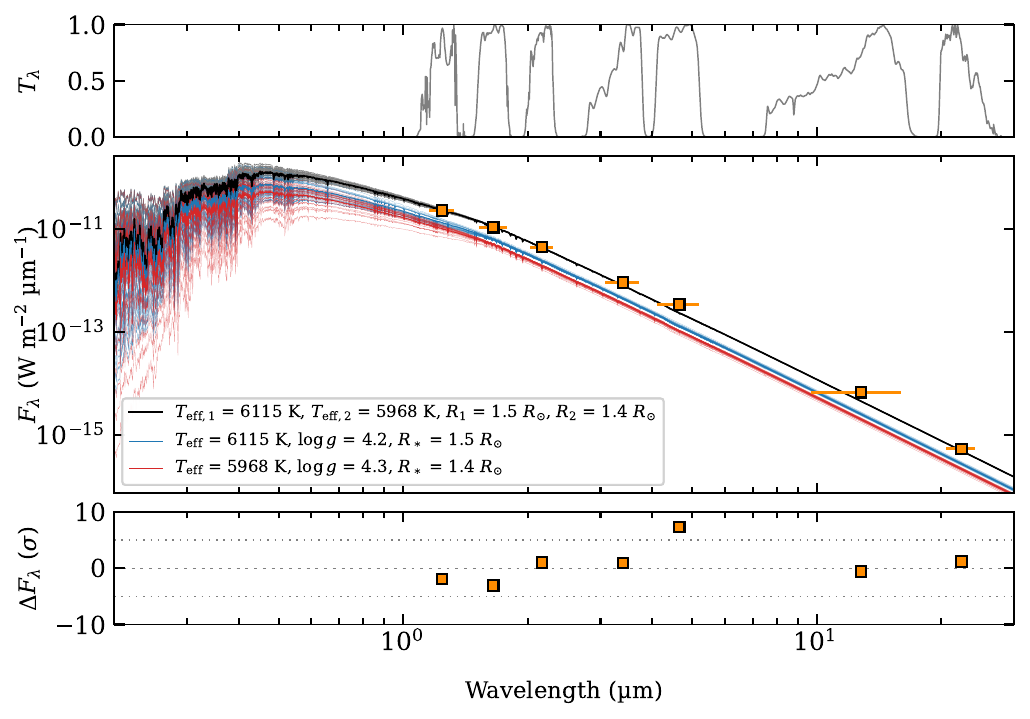}
        \caption{BT-NextGen stellar model atmospheres fitted to archival spectrophotometry (orange and green data points) of YSES~2~A (top), HD~91881~AB (centre), and HD~123227~AB (bottom).
        For the two binary stars, the SEDs for the binary A and B components are shown in blue and red, respectively.
        30 randomly drawn samples from the posterior are also shown with transparent curves.}
        \label{fig:stellar_sed}
    \end{centering}
\end{figure}

\vfill\eject

\section{Posterior distributions of the background fit}

\begin{figure}[h!]
    \begin{centering}
        \includegraphics[width=\columnwidth]{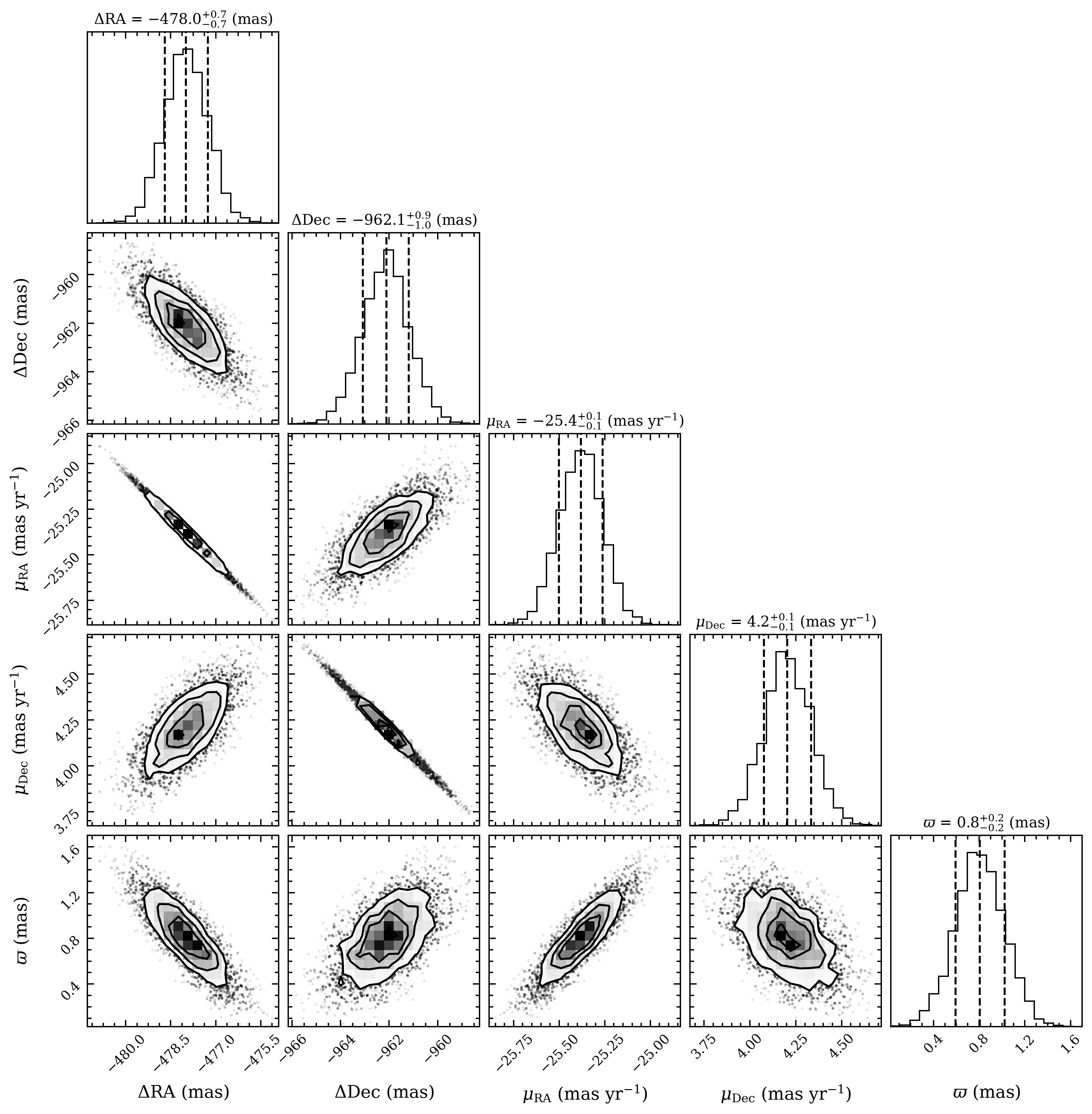}
        \caption{Posterior distribution from fitting the relative astrometry with the background model.
        The coordinates, RA and Dec, are given relative to the \emph{Gaia} DR3 coordinates of YSES\,2 at the J2016 epoch.}
        \label{fig:backtrack_corner}
    \end{centering}
\end{figure}

\vfill\eject

\onecolumn
\FloatBarrier

\section{GRAVITY Observations}
\label{app:gravity}

\begin{table*}[h!]
    \centering
    \caption{VLTI/GRAVITY observing log of YSES~2b.}
    \begin{tabular}{ccccccccc}
    Date & \multicolumn{2}{c}{Targets} & \multicolumn{2}{c}{NEXP/NDIT/DIT} & Airmass & $\tau_0$ & Seeing & $\gamma$ \\
    (UT) & SC & FT & B or b & A (if swap) & & (ms) & $^{\prime\prime}$ & \\
    \hline\hline
    2022-03-20 & HD~91881~A/B & HD~91881~B/A & 8/48/1~s & 8/48/1~s & 1.01-1.32 & 2.4-3.4~ms & $0.7-1.3^{\prime\prime}$  & ...  \\
    2022-03-20 & YSES~2~b & YSES~2~A & 9/8/100~s &  & 1.35-1.45 & 2.4-4.5~ms & $0.6-1.0^{\prime\prime}$  & 0.68 \\
    \hline
    2022-03-21 & HD~91881~A/B & HD~91881~B/A & 2/48/1~s & 2/48/1~s & 1.04-1.06 & 5.2-6.6~ms & $0.6-0.8^{\prime\prime}$  & ...  \\
    2022-03-21 & YSES~2~b & YSES~2~A & 3/8/100~s &  & 1.35-1.45 & 5.1-7.1~ms & $0.5-0.7^{\prime\prime}$ & 0.68 \\
    \hline
    2023-05-10 & HD~91881~A/B & HD~91881~B/A & 2/96/0.5~s & 2/96/0.5~s & 1.14-1.16 & 7.8-9.9~ms & $0.7-0.8^{\prime\prime}$  & ...  \\
    2023-05-10 & YSES~2~b & YSES~2~A & 8/4/100~s &  & 1.40-1.53 & 8.9-12.2~ms & $0.5-0.6^{\prime\prime}$ & ? \\
    2023-05-10 & YSES~2~A & YSES~2~A & & 2/16/10~s  & 1.57-1.59 & 8.3-11.3~ms & $0.4-0.6^{\prime\prime}$ & 0.99 \\
    \hline
    2023-06-03 & YSES~2~b & YSES~2~A & 4/4/100~s &  & 1.48-1.55 & 4.5-7.9~ms & $0.6-0.9^{\prime\prime}$ & ? \\
    2023-06-03 & HD~123227~A/B & HD~123227~B/A & 2/96/0.5~s & 2/96/0.5~s & 1.11-1.12 & 5.5-7.9~ms & $0.6-0.7^{\prime\prime}$  & ...  \\
    \hline
    \end{tabular}
    \label{tab:gravity_obs_log}
\end{table*}

\begin{figure*}[h!]
    \centering
    \includegraphics[width=\linewidth]{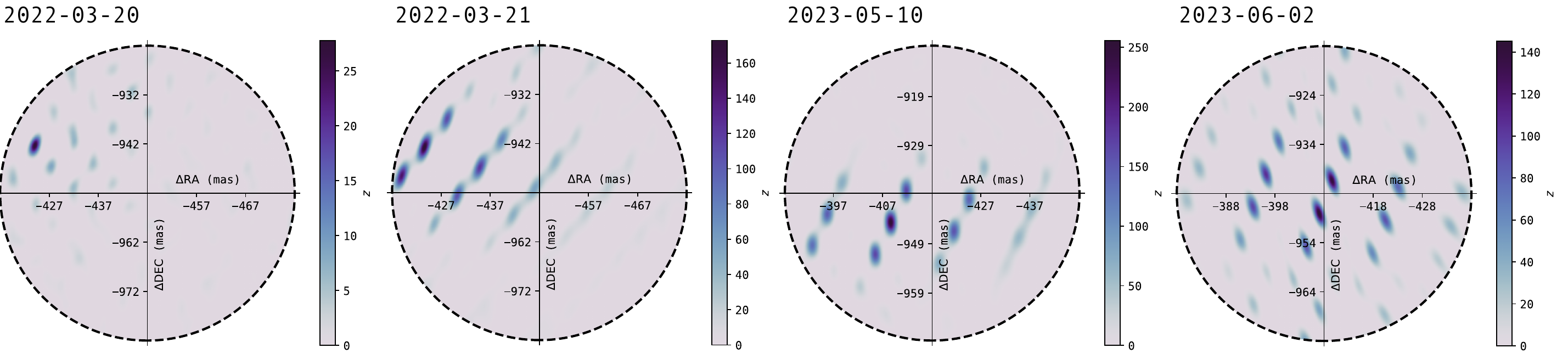}
    \caption{VLTI/GRAVITY detections of YSES~2b.}
    \label{fig:gravity_detections}
\end{figure*}

\end{appendix}

\end{document}